\documentclass{article}
 
\usepackage[final]{cgpg}

\input amssym.def
\input amssym.tex

\newtheorem{prop}{Proposition}[section]

\begin{document}

\title{Linearized general relativity and the Lanczos potential}

\author{Daniel Cartin\\
{\it Center for Gravitational Physics and Geometry, Department of Physics}\\
{\it Pennsylvania State University, University Park, PA 16802 USA}}

\date{24 October 1999}

\maketitle

\begin{abstract}

Recently, there has been a revival of interest in the Lanczos 
potential of the Weyl conformal tensor. Previous work by Novello and Neto
has been done with the linearized Lanczos potential as a model of a spin-2
field, which depends on a massless limit of the field. In this paper, we
look at an action based on a massless potential, and show that it is
classically equivalent to the linearized regime of general relativity,
without reference to any limiting case.

\end{abstract}

\section{Introduction}

It is well-known that the quantization of gravity has been a challenging
problem unsolved for several decades. There have been a number
of attempts to rewrite general relativity in terms of different variables
to provide an alternate viewpoint that might prove fruitful. In general,
these new ideas -- such as connection variables, twistors, and the null
surface formalism -- are based on geometric notions. However, we could
also follow the lead of Feynman~\cite{feyn}: start with a linear action
for a spin-2 field, and build up general relativity by requiring
consistency and gauge invariance. Certainly, in any tensor with more than
two indices, we can find a spin-2 piece to work with. Yet we can
combine these two approaches, since there exists a spin-2 piece of a
tensor (in addition to the metric) that has a geometric
interpretation -- namely, the Lanczos potential $L_{abc}$~\cite{lanczos}.
On a general spacetime, it has been shown~\cite{bam-cav, illge} that the
Weyl tensor can be written as\footnote{We use the notation \[ A_{(ab)} =
\frac{1}{2}
(A_{ab} + A_{ba}) \qquad A_{[ab]} = \frac{1}{2} (A_{ab} - A_{ba}) \] with
a comma denoting partial, and a semicolon covariant, differentiation.}
\begin{equation}\label{weyl-lanc}
C_{abcd} = 2 L_{ab[c;d]} + 2 L_{cd[a;b]}
- g_{a[c} (L_{|b|d]} + L_{d]b}) + g_{b[c} (L_{|a|d]} + L_{d]a}) +
\frac{2L}{3} g_{a[c} g_{d]b}
\end{equation}
where we have defined
\[
L_{ab} \equiv \tensor<L_a^c_b;c> - \tensor<L_a^c_c;b> = 2
\tensor<L^c_a[c;b]> \qquad L \equiv \tensor<L^a_a> = 2 \tensor<L^ab_[a;b]>
\]
Equation $(\ref{weyl-lanc})$ is analogous to the construction of the
Maxwell field from the vector potential. The Lanczos potential has the
symmetries
\[
L_{abc} = - L_{bac} \qquad L_{[abc]} = 0
\]
There are two additional properties that are usually assumed -- the {\it
algebraic gauge} $\tensor<L^{ab}_b> = 0$ and the {\it differential gauge}
$\tensor<L_{abc}^{;c}> = 0$. The proof of the relation $(\ref{weyl-lanc})$
given by Bampi and Caviglia does not depend on the choice of either
gauge\footnote{In most papers on the Lanczos potential, the word "gauge" is
used rather loosely to reflect the freedom to pick the values of the
tensors $\tensor<L^{ab}_b>$ and $\tensor<L_{abc}^{;c}>$. However, there is
a precise geometric meaning behind this, as shown by Hammon and
Norris~\cite{ham-nor}, namely, that we can think of $L_{abc}$
as a certain piece of a connection on an affine frame bundle over
spacetime with structure group $GL(4) \ltimes T^1_2 \Re^4$. The fact
that the Lanczos potential is related to the translations $T^1_2 \Re^4$
(and hence is Abelian)
gives rise to the fact that the Weyl tensor is linear in $L_{abc}$.}; the
spinorial proof of Illge assumes the algebraic gauge,
since in this case, the Lanczos potential is associated with a spinor
$L_{ABCC'} = L_{(ABC)C'}$, a natural choice for a spin-2 field. For the
rest of this paper, we will assume that both gauge conditions hold, so
that $L_{ab} = \tensor<L_a^c_b;c>$ and $L = 0$.

Now we can consider linear actions of the Lanczos potential as an
alternate model for gravity, which has been done previously by Novello and
Neto~\cite{nov-net1, nov-net2}. However, their work starts with an action
for a massive field, followed by a massless limit to recoup linearized
general relativity. This limit becomes problematic when one has to
consider the interaction of $L_{abc}$ with matter; as explained below, the
massless limit seems to imply that interactions in the Lanczos potential
picture correspond to no interaction in the related metric picture. The
main goal of this paper is to relate the two models of linearized general
relativity using perturbations of the Lanczos potential and the metric
without relying on a massless limit. By using some recent results of
Andersson and Edgar~\cite{and-edg}, we can derive the relation without any
extra conditions on the perturbations.

\section{Linear spin-2 fields}

On a flat background, we can consider the spacetime metric to be of the
form
\[
g_{ab} = \eta_{ab} + h_{ab}
\]
where $\eta_{ab}$ is the flat metric, and write out the form of the
conformal tensor $C_{abcd}$. When we compare this form to
that in terms of the Lanczos potential $(\ref{weyl-lanc})$, choosing the gauge
\begin{equation}\label{metric-gauge}
\tensor<h^{ab}_{,b}> = \frac{1}{2} h^{,a}
\end{equation}
with $h = \tensor<h^a_a>$, and using the tensor $\eta^{ab}$ to raise and
lower indices, we have that~\cite{lanczos}
\begin{equation}\label{lin-lanczos}
L_{abc} = \frac{1}{2} \biggl( h_{c[a,b]} - \frac{1}{6} \eta_{c[a} h_{,b]}
\biggr)
\end{equation}
This expression\footnote{As an aside, we note that this relation
illuminates the discussion in Section 5.7 of Penrose and
Rindler~\cite{pen-rin}, and especially the origin of equations such as
(5.7.12) in that work.}, valid to linear order, is a special case of a more
general
theorem~\cite{and-edg}. Suppose we have a tensor with the symmetries of the
Lanczos potential; then, on a Ricci-flat spacetime with Lorentz signature 
metric $g_{ab}$, we can locally find a traceless tensor ${\tilde K}_{ab}$,
such that
\begin{equation}\label{decomp1}
L_{abc} = {\tilde K}_{c[a;b]} + \frac{1}{3} g_{c[a} \tensor<{\tilde 
K}_{b]d}^{;d}>
\end{equation}
(assuming ${\tilde K}_{ab}$ is real). Since ${\tilde K}_{ab}$ is 
traceless, we can write it as ${\tilde K}_{ab} = K_{ab} - \frac{1}{4} 
g_{ab} K$, where $K = g^{ab} K_{ab}$.  Then, $(\ref{decomp1})$ becomes
\begin{equation}\label{decomp2}
L_{abc} = K_{c[a;b]} - \frac{1}{3} g_{c[a} K_{;b]} + \frac{1}{3} 
g_{c[a} \tensor<K_{b]d}^{;d}>
\end{equation}
With $K_{ab} = \frac{1}{2} h_{ab}$ and using $(\ref{metric-gauge})$,
we arrive at $(\ref{lin-lanczos})$.

Now we briefly outline the work of Novello and Neto. By writing down an action
made of massive fields $h_{ab}$ and $L_{abc}$~\cite{nov-net2} (again using
the flat metric to raise indices),
\begin{equation}\label{NN-action}
S_1 = \int \biggl[ h^{ab}( \tensor<L_a^c_{b,c}> - Q \tensor<L_a^c_{c,b}>
- Q \eta_{ab} \tensor<L^{ab}_{a,b}> ) + \frac{1}{2} m^2 h^{ab} h_{ab}
+ \frac{1}{4} L^{abc} L_{abc} - \frac{B - Q}{2(1 - 3B)} \tensor<L^{ac}_c>
\tensor<L_a^b_b> \biggr]
\end{equation}
where $Q = \frac{1 - B}{1 - 3B}$ and B is arbitrary, we find that the equations of motion are
\begin{equation}\label{bridge1}
L_{abc} = 2 (h_{c[a,b]} - B \eta_{c[a} h_{,b]} + B \eta_{c[a}
\tensor<h_{b]d}^{,d}>)
\end{equation}
and
\begin{equation}\label{bridge2}
h_{ab} = - \frac{1}{m^2} \tensor<L_{(a}^c_{b),c}> + \frac{Q}{m^2}
\tensor<L_{(a}^c_{|c|,b)}> + \frac{1}{3m^2} \eta_{ab} \tensor<L^{cd}_{c,d}>
\end{equation}
Note that $(\ref{bridge1})$ has the same form as $(\ref{decomp2})$ when
$B = 1/3$. Because we now have expressions for each of the fields in
terms of
derivatives of the other, we can select either one to be an auxiliary
field, and substitute the appropriate formula into the action, giving
a Lagrangian for a single field. Solving for $L_{abc}$ in terms of
$h_{ab}$ gives the usual linearized Einstein-Hilbert action with
a mass term, while the reverse results in
\begin{equation}\label{WL-action}
S_{2} = \int \biggl[ \frac{1}{16} C^{abcd} C_{abcd} + \frac{1}{2} m^2
L^{abc} L_{abc} \biggr]
\end{equation}
where $C^{abcd}$ is given in terms of the linearized Lanczos 
potential $L_{abc}$ as in $(\ref{weyl-lanc})$. Varying with respect 
to the potential gives
\begin{equation}\label{lanczos-field}
\tensor<C^{abcd}_{,d}> + m^2 L^{abc} = 0
\end{equation}

We cannot take the massless limit here if we
started with $(\ref{NN-action})$, since then, we would be unable to get
the relation $(\ref{bridge2})$, and hence we could not solve for $h_{ab}$
in terms of $L_{abc}$ to arrive at $(\ref{WL-action})$. However, if we
take $(\ref{WL-action})$ as our starting point, then there is no
difficulty in taking a massless limit. The limit does cause problems when
one tries to add the interaction of $L_{abc}$ to matter. We can certainly
add a term of the form
\begin{equation}\label{matter-action}
S_{int} = k' \int L_{abc} J^{abc}
\end{equation}
but then the question arises of how to relate the matter current $J_{abc}$
to the usual energy-momentum tensor $T_{ab}$. With a massive $L_{abc}$, we
can use $(\ref{bridge2})$ and start with the interaction in the metric
picture to get
\begin{equation}\label{interaction}
S_{int} = k \int h_{ab} T^{ab} = \frac{k}{m^2} \int \biggl[ -
\tensor<L_{(a}^c_{b),c}> + Q \tensor<L_{(a}^c_{|c|,b)}> + \frac{1}{3}
\eta_{ab} \tensor<L^{cd}_{c,d}> \biggr] T^{ab}
\end{equation}
As we can see, there are hazards in taking the massless limit -- from
$(\ref{interaction})$, it seems that taking the limit such that $k / m^2$
remains finite is equivalent to the limit $k \to 0$, i.e. in the metric
picture, there is no interaction between matter and the
graviton\footnote{In another paper~\cite{nov-net1}, the starting point of
Novello and Neto is $(\ref{WL-action})$, but varying both the metric and
the Lanczos potential. However, since an equation like $(\ref{bridge2})$
has to be used to get rid of the dependence on the metric, then the same
problem results.}. To avoid this, we start from a different viewpoint. We
can certainly write down the action $(\ref{WL-action})$ for a spin-2 field
and find its equations of motion with $m = 0$. The question is how we can
relate the solutions $L_{abc}$ with the solutions $h_{ab}$ of the
linearized Einstein equations. Because of the theorem of Andersson and
Edgar, we will always be able locally to find a symmetric tensor $h_{ab}$
such that its derivatives, when combined in the form $(\ref{decomp2})$,
give a massless solution $L_{abc}$. From
these equations of motion $(\ref{lanczos-field})$, it is natural to say
that $L_{abc}$ is a perturbation of the Lanczos potential, implying
$h_{ab}$ is a perturbation of the metric. Then we have the following
result.

\begin{prop}
Every solution $L_{abc}$ of $(\ref{lanczos-field})$ with $m=0$ corresponds to a solution 
$h_{ab}$ of the linearized Einstein-Hilbert action, up to gauge 
transformations of $h_{ab}$.
\end{prop}

\begin{proof}
To show that $h_{ab}$ gives rise to a solution $L_{abc}$, we 
use $(\ref{lin-lanczos})$ to form the linearized Lanczos 
potential in the gauge $(\ref{metric-gauge})$.  To go in the opposite
direction, suppose that $h_{ab}$ and 
$h'_{ab}$ are two metric perturbations that give the same Lanczos 
potential $L_{abc}$.  Then, $\Delta h_{ab} \equiv h'_{ab} - h_{ab}$ 
gives a Lanczos potential of zero (implying the Weyl tensor is zero)
and satisfies the field equations 
$R_{ab}(\Delta h_{cd}) = 0$.  Because of this, the Riemann tensor 
constructed from $\Delta h_{ab}$ is identically zero; hence, $\Delta 
h_{ab}$ corresponds to a constant multiple of a flat metric to linear 
order, which can be absorbed by a gauge transformation.
\end{proof}

To see the relation between the two models with the addition of matter, we
can introduce a coupling to a matter current of the form
$(\ref{matter-action})$. Using ideas similar to those above, we find the
relation between $J_{abc}$ and the more usual matter tensor $T_{ab}$
without refence to the mass. Using $(\ref{decomp2})$ to write the current
in terms of a symmetric tensor $T_{ab}$, we will show that $T_{ab}$ acts
like as a energy-momentum tensor of matter. Assuming that $L_{abc}$ is in
the algebraic gauge, then $(\ref{matter-action})$ is invariant under the
transformation $J^{abc} \to J^{abc} + \eta^{c[a} V^{b]}$. Thus, by
choosing $V^a = - \frac{1}{3} \tensor<T^{ad}_{,d}>$, we have that

\[
J^{abc} = T^{c[a,b]} - \frac{1}{3} \eta^{c[a} T^{,b]}
\]
and the equations of motion become
\begin{equation}\label{current}
\tensor<C^{abcd}_{,d}> = J^{abc} = T^{c[a,b]} - \frac{1}{3} \eta^{c[a} 
T^{,b]}
\end{equation}
These equations, along with the Einstein equations and the Bianchi
identities, written solely in terms of the metric for the Weyl tensor,
\[
\tensor<C^{abcd}_{,d}> = R^{c[a,b]} - \frac{1}{6} \eta^{c[a} R^{,b]}
\]
lead to the
identification of the tensor $T^{ab}$ as the energy-momentum tensor.
The fact that the divergence of $T^{ab}$ is zero comes from the
trace-free nature of the left-hand side of $(\ref{current})$:
\begin{eqnarray*}
\tensor<C^{ab}_b^d_{,d}> &=& 0 \ =\  \tensor<J^{ab}_b> \\
&=& \frac{1}{2} \biggl[ \tensor<T^{ab}_{,b}> - T^{,a} - \frac{1}{3} T^{,a}
+ \frac{4}{3} T^{,a} \biggr] \\
&=& \frac{1}{2} \tensor<T^{ab}_{,b}>
\end{eqnarray*}

If we start with the same action $(\ref{WL-action})$ on a curved 
background, we find that the equations of motion are inconsistent 
unless the Weyl tensor of the background spacetime is zero. 
However, this is not a disability, but rather a sign that, if we 
consider the gravitational degrees of freedom to reside in the 
conformal curvature, then it does not make sense to start with a 
background that already has gravitational interactions.

\section{Discussion}

In this paper, we have shown that there is a relationship between 
linearized general relativity written in terms of perturbations of the
metric and the 
Lanczos potential, when expanded on a flat background 
spacetime. Unlike in~\cite{nov-net1, nov-net2}, there 
is no need to solve for bridging formulas between the two fields using the 
equations of motion, so this relationship holds regardless of the mass
of the field. Because our action in written only with the linearized
Lanczos potential, one can hope that the full non-linear theory can be more
easily arrived at. As a sketch of the procedure, we note that there are four
curvature invariants that can be formed out of the Weyl tensor, namely,
\[
\tensor<C_{abcd}> C^{abcd}, \quad \tensor^*<C_abcd> C^{abcd}, \quad C_{abcd}\
\tensor<C^cd_ef> C^{abef}, \quad \tensor^*<C_ab^cd> \tensor<C_cd^ef>
\tensor<C_ef^ab>
\]
(The quantity $\tensor^*<C_abcd>$ is the dual of the Weyl tensor, 
$\tensor^*<C_abcd> = \frac{1}{2} e_{abef} \tensor<C^{ef}_{cd}>$). Thus, 
any analytic Lagrangian must be written as a polynomial in these four 
quantities. Because we can extend $(\ref{lin-lanczos})$ to arbitrary
order in $h_{ab}$, by substitution, one can relate the equations of motion
of such an
action to the Einstein equations. Another, perhaps more elegant, method is to
use the ideas of Deser~\cite{deser}, deriving an action to all orders from the
requirements of consistency and gauge invariance. We hope to return to this
problem in the future.

\section{Acknowledgements}

The author would like to thank Sameer Gupta, Gaurav Khanna and Lee Smolin for many helpful comments, and
was supported in part by NSF grant PHY-951420 to the Pennsylvania State University and a gift from the
Jesse Phillips Foundation.

\end{document}